\documentclass[sigconf,balance]{acmart}
\AtBeginDocument{%
  }

\setcopyright{acmlicensed}
\copyrightyear{2018}
\acmYear{2018}
\acmDOI{XXXXXXX.XXXXXXX}

\acmConference[Conference acronym 'XX]{Make sure to enter the correct
  conference title from your rights confirmation email}{June 03--05,
  2018}{Woodstock, NY}

\acmISBN{978-1-4503-XXXX-X/2018/06}

\usepackage[utf8]{inputenc} 
\usepackage[T1]{fontenc}    
\usepackage{hyperref}       
\usepackage{url}            
\usepackage{booktabs}       
\usepackage{amsfonts}       
\usepackage{nicefrac}       
\usepackage{microtype}      
\usepackage{xcolor}         
\usepackage{algpseudocode}
\usepackage{lipsum}
\usepackage{xspace}
\usepackage{fancyvrb}
\usepackage{graphicx}
\usepackage{caption}
\usepackage{soul}
\usepackage{adjustbox}
\usepackage{multirow}
\usepackage{comment}
\usepackage[dvipsnames]{xcolor}
\usepackage{wrapfig}
\usepackage{enumitem}
\usepackage{tabularx}
\usepackage{float}
\usepackage{amsmath}
\usepackage{cleveref}
\usepackage{stmaryrd} 
\usepackage{balance} 
\usepackage[table,xcdraw,dvipsnames]{xcolor}
\usepackage[many]{tcolorbox}
\usepackage{makecell}
\usepackage{bm}
\usepackage{algorithm}
\usepackage{algpseudocode}
\usepackage{siunitx}
\usepackage{booktabs,makecell}
\usepackage{pifont}
\algrenewcommand\algorithmicrequire{\textbf{Input:}}
\algrenewcommand\algorithmicensure{\textbf{Output:}}
\algrenewcommand{\algorithmiccomment}[1]{\hspace{.5em}\(\triangleright\)\,#1}
\algtext*{EndIf}
\algtext*{EndFor}
\algtext*{EndWhile}

\usepackage{xcolor}

\newcommand{\reddown}{\textcolor{red}{\downarrow}}

\newcolumntype{R}[2]{%
    >{\adjustbox{angle=#1,lap=\width-(#2)}\bgroup}%
    l%
    <{\egroup}%
}

\newcommand{\crux}{CRUXEval\xspace}
\newcommand{\heval}{HumanEval\xspace}
\newcommand{\avatar}{Avatar\xspace}
\newcommand{\ceval}{ClassEval\xspace}
\newcommand{\codemind}{CodeMind\xspace}
\newcommand{\reval}{REval\xspace}

\newcommand{\name}{\textsc{R$^2$Eval}\xspace}

\newcommand{\dsr}{DeepSeek-R1\xspace}
\newcommand{\gptf}{GPT-4.1\xspace}
\newcommand{\gemini}{Gemini\xspace}

\newcommand{\omini}{o4-mini\xspace}
\newcommand{\dsc}{DeepSeek-Coder\xspace}

\definecolor{myhighlight}{HTML}{FCFF2F}

\copyrightyear{2026}
\acmYear{2026}
\setcopyright{cc}
\setcctype{by}
\acmConference[ICSE-Companion '26]{2026 IEEE/ACM 48th International Conference on Software Engineering}{April 12--18, 2026}{Rio de Janeiro, Brazil}
\acmBooktitle{2026 IEEE/ACM 48th International Conference on Software Engineering (ICSE-Companion '26), April 12--18, 2026, Rio de Janeiro, Brazil}
\acmPrice{}
\acmDOI{10.1145/3774748.3787747}
\acmISBN{979-8-4007-2296-7/2026/04}

\begin{document}

\title{Evaluating LLMs Code Reasoning Under Real-World Context}

\author{Changshu Liu}
\email{cl144@illinois.edu}
\affiliation{%
  \institution{University of Illinois Urbana-Champaign}
  \city{Urbana}
  \state{Illinois}
  \country{USA}
}

\begin{abstract}
Code reasoning tasks are increasingly crucial to evaluating large language models (LLMs). Yet most existing benchmarks rely on simplistic, LLM-generated snippets or human-written solutions to code challenges and often restrict inputs and outputs to primitive types, failing to reflect the structure and dependencies of real-world projects. These simplifications limit their ability to measure practical generalizability. We present \name\footnote{\name stands for \underline{R}ealistic Code \underline{R}easoning \underline{Eval}uation}, a benchmark of $135$ code reasoning problems drawn from ten widely used Python projects. Unlike prior work, \name serializes compound and custom types, preserving real-world data complexity and enabling a more realistic assessment of LLMs.
\end{abstract}


\maketitle

\begin{figure*}
    \centering
    \includegraphics[width=0.88\linewidth]{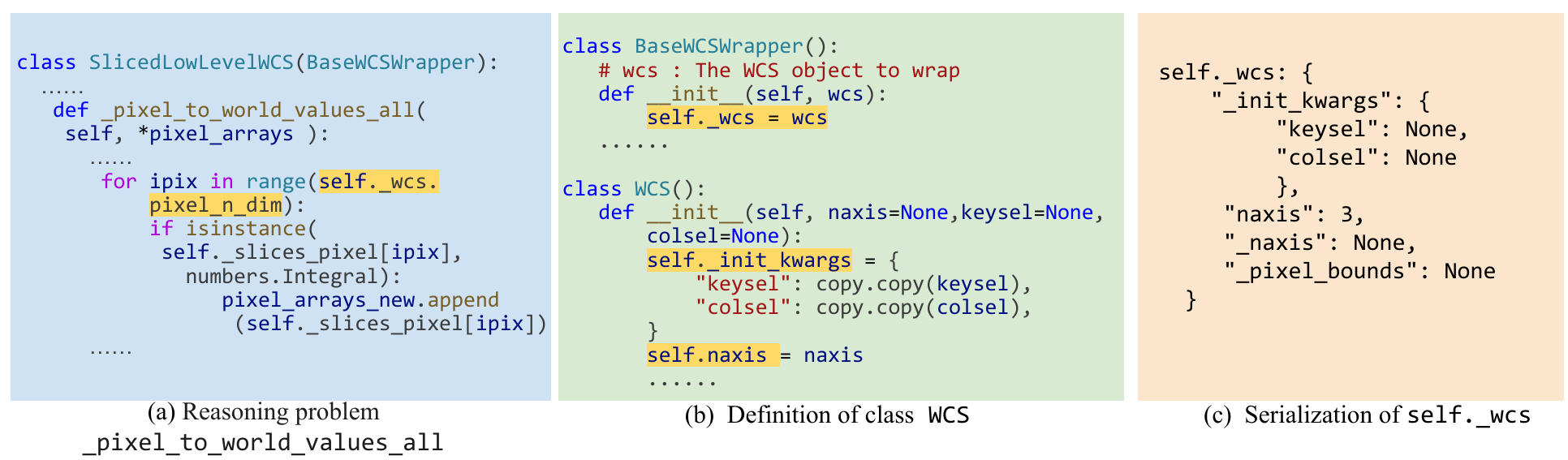}
    \caption{Example of custom type variable serialization in \name }
    \label{fig:ser}
    \vspace{-5pt}
\end{figure*}

\section{Research Problem \& Motivation}
Recently, code reasoning, which measures LLMs’ capability of understanding the behavior
of programs during execution, has become a popular evaluation strategy to assess LLMs.
This capability is particularly crucial for tasks that require the simulation of the control/data flow of the program, e.g., code translation\cite{ibrahimzada2025alphatrans, pan2024lost, ibrahimzada2024program}, program repair~\cite{tang2025co, xia2024agentless}, and code generation~\cite{liu2023your,jiang2024survey, ding2024semcoder}.

Existing benchmarks assess LLMs' code reasoning ability in input/output prediction on either LLM-generated snippets or human-written solutions to programming challenges, but neither captures the structural and contextual complexity of production code. \crux~\cite{gu2024cruxeval} contains short, self-contained Python programs. Datasets such as \avatar~\cite{ahmad2023avatar}, \ceval~\cite{du2023classeval}, and \heval~\cite{chen2021evaluating} are more challenging—some problems involve code spanning multiple methods—yet they remain far from reflecting the rich inter- and intra-procedural dependencies typical of real repositories. CodeSense~\cite{roy2025codesense} makes a step further and extracts methods from real-world projects, but it discards those with non-primitive variables in the input/output.
Consequently, although state-of-the-art LLMs perform strongly on the existing benchmarks, especially the widely used \crux, their ability to generalize to real-world projects remains uncertain and warrants further investigation. 

\vspace{-8pt}
\section{Approach and Contributions}
To overcome the limitation of existing approaches, we propose \name, a code reasoning benchmark drawn from real-world projects. Our pipeline combines program analysis to automatically serialize/deserialize compound, complex, and custom types, well beyond the primitive-only assumptions common in prior work \cite{kopelexecution, roy2025codesense}. 

\sloppy \name first collects code reasoning problems from ten popular Python projects, including: \texttt{scikit-learn}, \texttt{django}, \texttt{requests}, \texttt{seaborn}, \texttt{sphinx}, \texttt{pytest}, \texttt{astropy}, \texttt{xarray}, \texttt{matplotlib}, and \texttt{sympy}.
\name extracts target methods from them and records the runtime inputs and outputs. A \textbf{big challenge} here is that values in real-world projects often involve complex, custom objects.
Unlike primitive types, e.g., \texttt{int}, \texttt{str}, which can be directly printed, custom objects often lack the actual implementation of \texttt{\_\_str\_\_()} or \texttt{str()}. Therefore, \textbf{printing such objects typically yields only their memory address or hash value, rather than actual values.}
To overcome this challenge, \name leverages static and dynamic program analysis to iteratively decompose complex data types of inputs/outputs into primitives (or core compound types), serializing them into JSON formats. The serialized inputs or outputs will be provided to the LLM, along with code and intra/inter-dependencies, to predict outputs or inputs. \textbf{Another big challenge} is \textbf{automated detection of false negatives}, which may happen if we textually compare predictions and ground truth. To overcome this in a scalable fashion, \name deserializes LLM's predictions and creates tests with objects to evaluate correctness at runtime. Figure~\ref{fig:ser}-a shows a reasoning problem \texttt{\_pixel\_to\_world\_values\_all} that takes a custom-typed variable \texttt{\_wcs} extended from the base class \texttt{BaseWCSWrapper} (Figure~\ref{fig:ser}-b) as input. \name serializes \texttt{\_wcs} into the JSON representation shown in Figure~\ref{fig:ser}-c. 

\name currently offers $135$ \textbf{real-world} reasoning problems, each represented as a triplet $\{P, I, O\}$, where $P$ is the code for the reasoning task (including relevant method dependencies and class context), and $I$/$O$ denote serialized method input(s)/output. Experiment results on \name show that the performance of assessed LLMs \textbf{significantly drops} by $64.32\%$ and $52.22\%$, on input prediction and output prediction, respectively, compared to their performance on \crux. These results \emph{question bold claims about code reasoning abilities of LLMs}, promote more \emph{meaningful evaluation of code reasoning}, aiming to \emph{advance LLMs for robust code reasoning under real-world settings}. 

\begin{figure*}
    \centering
    \includegraphics[width=0.90\linewidth]{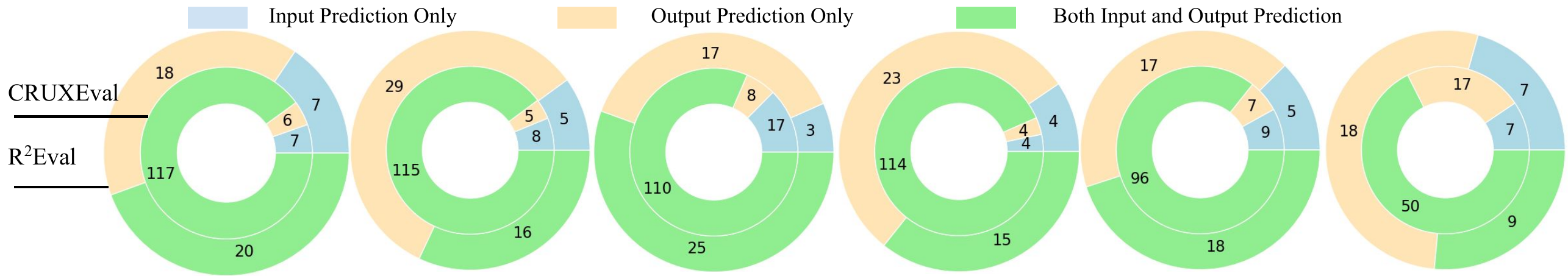}
    \caption{Unique and common problems each LLM succeeds in predicting their inputs and outputs}
    \label{fig:overlap}
\end{figure*}

\begin{table*}[]
\caption{Code Reasoning Performance of LLMs on \name and \crux. The $\reddown$ indicates the performance drop from \crux to \name}
\label{tab:results}
\vspace{-10pt}
\setlength{\tabcolsep}{6pt}
\scalebox{0.90}{
\begin{tabular}{|l|cc|cc|}
\hline
\multirow{2}{*}{\textbf{Subject LLMs}} &
  \multicolumn{2}{c|}{\textbf{\name}} &
  \multicolumn{2}{c|}{\textbf{CRUXEval}} \\ \cline{2-5} 
 &
  \multicolumn{1}{c|}{\textbf{Input Prediction(\%)}} &
  \textbf{Output Prediction(\%)} &
  \multicolumn{1}{c|}{\textbf{Input Prediction(\%)}} &
  \textbf{Output Prediction(\%)} \\ \hline
\textbf{o4-mini}                & \multicolumn{1}{c|}{20.00 ($\reddown$ \textcolor{red}{72.59})}    & 28.15 ($\reddown$ \textcolor{red}{63.70}) & \multicolumn{1}{c|}{92.59} & \textbf{91.85} \\ \hline
\textbf{Gemini-2.5-Pro}         & \multicolumn{1}{c|}{15.56 ($\reddown$ \textcolor{red}{76.29})} & \textbf{34.07} ($\reddown$ \textcolor{red}{54.82}) & \multicolumn{1}{c|}{91.85} & 88.89 \\ \hline
\textbf{DeepSeek-R1}            & \multicolumn{1}{c|}{\textbf{21.48} ($\reddown$ \textcolor{red}{73.33})} & 31.85 ($\reddown$ \textcolor{red}{55.56})& \multicolumn{1}{c|}{\textbf{94.81}} & 87.41 \\ \hline
\textbf{GPT-4.1}                & \multicolumn{1}{c|}{14.81 ($\reddown$ \textcolor{red}{72.60})} & 28.15 ($\reddown$ \textcolor{red}{59.26}) & \multicolumn{1}{c|}{87.41} & 87.41 \\ \hline
\textbf{Gemini-1.5-Pro}         & \multicolumn{1}{c|}{17.04 ($\reddown$ \textcolor{red}{60.74})} & 25.93 ($\reddown$ \textcolor{red}{50.37})& \multicolumn{1}{c|}{77.78} & 76.30  \\ \hline
\textbf{DeepSeekCoder-33B-Inst} & \multicolumn{1}{c|}{12.59 ($\reddown$ \textcolor{red}{30.37})} & 20.74 ($\reddown$ \textcolor{red}{29.63}) & \multicolumn{1}{c|}{42.96} & 50.37 \\ \hline
\textbf{Total}                  & \multicolumn{1}{c|}{16.91 ($\reddown$ \textcolor{red}{64.32})} & 28.15 ($\reddown$ \textcolor{red}{52.22}) & \multicolumn{1}{c|}{81.23} & 80.37 \\ \hline
\end{tabular}
}
\vspace{-5pt}
\end{table*}

\section{Evaluation}
We compare \name with \crux, a widely used benchmark for code reasoning. Because \name is smaller, we sample $135$ problems from \crux to match \name’s size. We evaluate six LLMs—\omini~\cite{o4-mini2025}, \gemini-2.5-Pro~\cite{gemini25pro2025}, \dsr~\cite{guo2025deepseek}, \gptf~\cite{openai2025gpt41}, \gemini-1.5-Pro~\cite{team2023gemini}, and \dsc-Inst-33B~\cite{guo2024deepseek}—and report input-prediction and output-prediction performance in Table~\ref{tab:results}. We have the following key findings:

\vspace{-5pt}
\begin{itemize} [leftmargin=*]
\item Reasoning LLMs (\omini, \gemini-2.5, and \dsr) consistently outperform non-reasoning LLMs across both \name and \crux. On average, the margin is $13.95\%$ for input prediction and $12.22\%$ for output prediction.

 \item Moving from \crux to \name, LLMs' performance drops substantially: \emph{$64.32\%$} for input prediction and \emph{$52.22\%$} for output prediction. We attribute this to code properties that are intrinsic to real-world programs but largely absent from \crux, including inter- and intra-procedural dependencies, third-party API usage, and non-primitive (compound or custom) data types.

\item Figure~\ref{fig:overlap} shows the unique and common problems each LLM succeeds in predicting their inputs and outputs on both \name and \crux. We observe that moving from \crux(inner doughnuts) to \name(outer doughnuts), the percentages of overlap
between input and output prediction success decrease drastically. We speculate that it is because in real-world projects, the complex code constructs and rich dependencies make the backward reasoning in the input prediction obviously more difficult than the forward reasoning in the output prediction.

\end{itemize}


\vspace{-9pt}
\section{Background and Related Work}
\crux~\cite{gu2024cruxeval} and \codemind~\cite{liu2024codemind} are early works in evaluating LLMs for input-ouptut prediction. \reval~\cite{chen2024runtime} and CES ~\cite{liu2025assessing} take one step forward and complement output prediction with prediction on the intermediate state of the code execution. 
ExerScope~\cite{liu2025tool} use program analysis tools to identify code properties impacting the performance of LLMs on code reasoning.
However, all these techniques collect reasoning problems from human-written/LLM-generated simple code that is far from real-world complexity. Another two closely related works are EXE~\cite{kopelexecution} and CodeSense~\cite{roy2025codesense}, which evaluate LLM code reasoning on open-source Python repositories; however, neither of them considers non-primitive data types in the input. In contrast, our work preserves the complexity of the input in the real-world project via serializing complex custom types into a readable JSON-like format. 


\bibliographystyle{ACM-Reference-Format}
\bibliography{sample-base}

\appendix

\end{document}